\documentclass[a4paper]{jpconf}
\usepackage{graphicx}\usepackage{epsfig,cite}\usepackage{graphicx, subfigure}
\begin{document}
\title{Perturbative corrections to $\bar B \rightarrow X_s \gamma$  in supersymmetry  at next-to-leading order~\footnote{ Based on 
an invited plenary talk  given by TH at the International Conference  on the Structure and Interactions of the Photon, PHOTON 2011,  Spa, Belgium, 22nd - 27th of May 2011, MZ-TH/11-36.}}

\author{Tobias Hurth}
\address{Institute for Physics,  Johannes Gutenberg University,\\D-55099 Mainz, Germany}
\author{Christoph Greub, Volker Pilipp, Christof Sch\"upbach}
\address{Albert Einstein Center for Fundamental Physics, Institute for Theoretical Physics,\\  CH-3012 Bern, Switzerland}
%\author{Matthias Steinhauser}
%\address{Institute for Theoretical Particle Physics, Karlsruhe Institute of Technology (KIT), \\D-76128 Karlsruhe, Germany}

\begin{abstract}
We give a brief overview about perturbative corrections to the inclusive decay mode 
 $\bar B \to X_s \gamma$  in supersymmetric models.
\end{abstract}

\section{Introduction}

Perturbative QCD corrections are well-known for being the dominant contributions to 
the radiative inclusive penguin decay~\cite{Hurth:2010tk,Hurth:2007xa,Hurth:2003vb}.
This perturbative dominance was recently reassured by a dedicated analysis~\cite{Benzke:2010js}
in which non-perturbative  corrections to the inclusive decay mode $\bar B \rightarrow X_s \gamma$
have been estimated to be well below $10\%$. 

Within a global effort,  a  perturbative QCD calculation to the
next-to-next-to-leading-logarithmic (NNLL) level within the Standard Model  (SM) 
has been
performed and has led to the first NNLL prediction of the $\bar B \to X_s  \gamma$ branching 
fraction~\cite{Misiak:2006zs}. Using the photon energy cut $E_0=1.6$ GeV,  the branching
ratio reads 
\begin{equation}\label{final1}
{\cal B}(\bar B \to X_s \gamma)_{\rm NNLL} =  (3.15 \pm 0.23) \times 10^{-4}.
\end{equation}
This result is based on various highly-nontrivial perturbative
calculations ~\cite{Misiak:2004ew,Bobeth:1999mk,Gorbahn:2004my,Gorbahn:2005sa,Czakon:2006ss,Blokland:2005uk,Melnikov:2005bx,Asatrian:2006ph,Asatrian:2006sm,Bieri:2003ue,Misiak:2006ab}. 
The combined experimental data according to the Heavy Flavor  Averaging Group
(HFAG)~\cite{hfag} leads to 
\begin{equation}
 {\cal B}(\bar B \rightarrow X_s \gamma) = (3.55 \pm 0.24 \pm 0.09) \times 10^{-4}
 \,,
\end{equation}
where the first error is combined statistical and systematic, and the second
is due to the extrapolation in the photon energy.  
Thus, the SM prediction and the experimental average are consistent at the $1.2 \sigma$ level.

This  is just one example among the impressive confirmation of the SM in all experiments 
in flavour physics during the last decade~\cite{Antonelli:2009ws, Buchalla:2008jp}, 
including   the first generation of the $B$ factories at KEK (Belle
experiment at the KEKB $e^+e^-$ collider)~\cite{Belle} and at SLAC (BaBar
experiment at the PEP-II $e^+e^-$ collider)~\cite{Babar},
and the Tevatron $B$ physics programs (CDF~\cite{TevatronB1} and
D0~\cite{TevatronB2} experiments). 
Also the first results of the LHCb experiment~\cite{LHCb} are in full agreement with the 
simple CKM theory of the SM.

This feature is somehow unexpected because in principle 
flavour changing neutral current (FCNC) processes  like  $\bar B \to X_s  \gamma$ 
offer high sensitivity to new physics (NP).
  Additional contributions to the
decay rate, in which SM particles in the loops are replaced by new particles such as
the supersymmetric charginos or gluinos are not suppressed by the loop 
factor $\alpha/4\pi$ relative to the SM contribution. Thus,  FCNC
decays provide information about the SM and its extensions via virtual
effects to scales presently not accessible otherwise. This approach is complementary to the
direct production of new particles at collider experiments.

\section{Supersymmetric flavour problem}

The experimental fact that none of the dedicated flavour experiments 
has observed any unambiguous sign
of new physics yet,  in particular  no ${\cal O}(1)$ NP effects in any FCNC 
process,  implies
the famous flavour problem, namely why FCNC processes are suppressed. 
It has to be solved in any viable new physics model. The hypothesis of minimal flavour 
violation (MFV)~\cite{Chivukula:1987py,Hall:1990ac,D'Ambrosio:2002ex}, 
i.e.\ that the NP model
has no flavour structures beyond the Yukawa couplings, solves the problem formally.
However,   new flavour structures beyond the Yukawa couplings are still compatible 
with the present data~\cite{Hurth:2009ke}   because the flavour sector has been tested only 
at the $10\%$  level  in the $b\to s$  transitions.

Today supersymmetric models are  often given priority in our search for NP
  beyond the SM. This is primarily suggested by theoretical 
arguments related to the well-known hierarchy problem. Supersymmetry 
eliminates the sensitivity for the highest scale in the theory and, thus, 
stabilizes the low energy theory.   There are other features in supersymmetric 
theories which are promising like the unification of the gauge couplings and the 
existence of  a dark matter candidate. Supersymmetry also represents the unique 
extension of Poincare symmetry.  

The precise mechanism of the necessary supersymmetry breaking is unknown. A 
reasonable approach to this problem is the inclusion of the most general 
soft breaking term consistent with the SM gauge symmetries in the 
so-called unconstrained minimal supersymmetric standard model
(MSSM). This leads to a proliferation of free parameters in the theory. 

The decay  $\bar B \rightarrow X_s \gamma$ is sensitive to the 
mechanism of supersymmetry breaking because,  
in the limit of exact supersymmetry, the decay rate would
be just zero:
\begin{equation}
{\cal B} (\bar B \to X_s \gamma)_{Exact\, Susy} = 0.  
\end{equation}
This follows from an argument first given by Ferrara and Remiddi in 1974
\cite{Ferrara}. 
In that work  the absence of the anomalous 
magnetic moment in a supersymmetric abelian gauge theory was shown.

In the MSSM there are new sources of  FCNC transitions. 
Besides  the CKM-induced contributions,
which are brought about by  a charged Higgs or a chargino,  
there are  generic supersymmetric  contributions that arise from  
flavour mixing  in the squark mass matrices in case they are not aligned 
to the ones in the quark sector. Then the gluino contribution enhanced
by an extra factor $\alpha_s$ instead of $\alpha_{\rm weak}$ significantly contributes 
to the decay rate.

Thus,  the general structure of the MSSM does not explain 
the suppression of 
FCNC processes, which is observed in experiments; the gauge symmetry
within the supersymmetric framework does not protect
 the observed strong suppression of the FCNC transitions. 
This is the crucial point of the well-known supersymmetric 
flavour problem.

\section{Parameter bounds from the inclusive decay $\bar B \rightarrow X_s \gamma$}

Parameter bounds on NP from flavour physics is  a model-dependent issue.
The present data on $\bar B \to X_s  \gamma$ implies a very stringent bound 
for example  on the inverse
compactification radius of the minimal universal extra dimension model
(mACD) ($1/R > 600 {\rm GeV}$ at $95\%$ CL)~\cite{Haisch:2007vb}.
The bound is  much stronger than the ones derived from other
measurements.  
Moreover, there is a bound induced by $\bar B \rightarrow X_s \gamma$ 
 on  the charged Higgs mass in the two Higgs-doublet model (II):  $M_{H^+} > 295{\rm GeV}$
at $95\%$ CL~\cite{Misiak:2006zs}. It  is based on a NLL QCD calculation within this model presented in Refs.~\cite{Ciuchini:1997xe,Borzumati:1998tg}.
The latter bound is not valid  in  general two-Higgs doublet models, especially in supersymmetric models. However, 
the two-Higgs-doublet model (II)  is  a good approximation for 
gauge-mediated  supersymmetric models with large $\tan\beta$, where the charged Higgs
contribution dominates the other supersymmetric  contributions.  

Simplifying assumptions about the  parameters
often introduce model-dependent correlations between different observables.  
Thus, flavour physics will also help in discriminating between  
the various  models that will be proposed by then.
In view of this, it is important to calculate the rate of the rare $B$ decays,
 with theoretical uncertainties as reduced 
as possible and general enough for generic 
supersymmetric models.

The rare decay $\bar B \rightarrow X_s \gamma$
has already carved out large  regions in the space of free
parameters of most of the supersymmetric models. 
Once more precise data from the Super $B$ factories are   available, 
this decay will undoubtedly gain even more efficiency
in selecting the viable regions of the parameter space in the various
classes of models. 
Constraints based on nontrivial QCD calculations within various supersymmetric extensions are heavily 
analyzed in the literature, see for example the Refs.\cite{Bertolini:1990if, Ciuchini:1998xy,Degrassi:2000qf,
  Carena:2000uj, Borzumati:1999qt, Besmer:2001cj,
  Ciuchini:2002uv, Ciuchini:2003rg, Okumura:2003hy, 
  Degrassi:2006eh, Ciuchini:2007ha,
  Altmannshofer:2008vr, Crivellin:2009ar,Crivellin:2011jt} .

Finally, model-independent analyses in the effective field theory
approach without~\cite{Ali:2002jg} and with the assumption of minimal
flavour violation~\cite{Hurth:2008jc} also show the
strong constraining power of the $\bar B \rightarrow  X_s \gamma$ branching
fraction.

\section{NLL calculations in supersymmetry}

While in the SM, the rate for $\bar B \to X_s \gamma$
is known up to NNLL in QCD, 
also within
supersymmetric theories higher order  calculations have been pushed forward in
recent years. 
At the LL level  there are several contributions to the decay amplitude: 
besides the contributions solely induced by flavour mixing in the quark sector 
with a $W$ boson or a charged Higgs boson and a top quark in the loop, there is
also a chargino contribution with an  up-type squark which can be induced by the CKM
matrix.  If we consider also generic new sources of flavour violation induced by a disalignement
of quarks and squarks, there are additional contributions from a chargino, gluino and also neutralino.
The first complete analysis of the decay rate of $\bar B \rightarrow X_s \gamma$ has been 
presented in Ref.~\cite{Bertolini:1990if}.

It is highly desirable to analyse   these  non-standard contributions   with NLL precision:  
Besides the large uncertainties in the LL predictions,
the step from the LL to the NLL precision is also necessary 
in order to check  the validity of the perturbative 
approach in the model under consideration. 
Moreover, it was already shown in specific NP  scenarios 
that bounds on the parameter space of  non-standard 
models are very sensitive to NLL contributions.

\subsection{NLL calculation in MFV}

The MFV hypothesis is a formal model-independent  solution to the 
NP flavour problem. It assumes that the flavour and the CP 
symmetry are  broken as in the SM.  Thus, it 
requires that all flavour- and CP-violating interactions be 
linked to the known structure of Yukawa couplings.  
A renormalization-group-invariant definition of MFV 
based on a symmetry principle 
is given in~\cite{D'Ambrosio:2002ex}; 
this  is mandatory for  a consistent  
effective field theoretical analysis
of NP  effects. 
The MFV hypothesis is an important benchmark. Because any measurement  which 
is inconsistent with the general constraints and relations induced by the MFV 
hypothesis~\cite{Hurth:2008jc}
  indicates the existence of new flavour structures.

This hypothesis  can also be used within the MSSM.  It can be implemented by
assuming that the squark and quark mass matrices can be simultaneously
diagonalized (alignement). 
In this case there are no flavour-changing interactions induced by the gluino 
at the tree level.

\begin{figure}
\begin{center}
\epsfig{figure=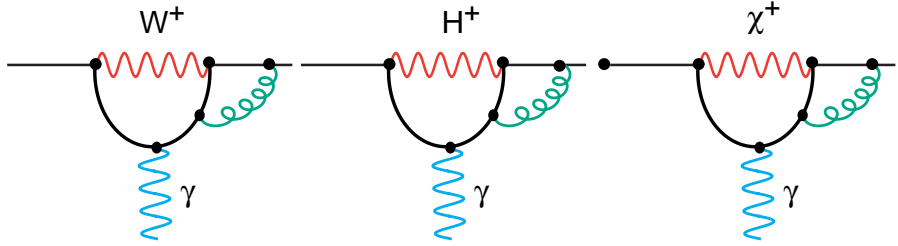,width=12cm}
\end{center}
\caption{Example diagrams for NLL gluonic corrections to the $W$ boson, charged Higgs and chargino contribution.}
\label{Figure1}
\end{figure}

The first NLL calculation of the inclusive decay $\bar B \rightarrow X_s \gamma$ in the MSSM 
with the MFV hypothesis includes the gluon corrections  to the charged Higgs and the chargino
contribution~\cite{Ciuchini:1998xy}, see Figure~\ref{Figure1}.  In particular the possibility of 
destructive interference of the chargino 
and the charged Higgs contribution is  studied. 
The analysis is   done under the MFV assumption 
that the only source of flavour violation  at the electroweak scale is that of the SM, 
encoded in the CKM matrix. Other flavour-changing interactions were  suppressed 
by assuming the gluino being heavy. 
It  is  found  
that,  in this  specific supersymmetric scenario,  
bounds on the parameter space are rather sensitive 
to NLL contributions 
and they lead to a significant reduction of the stop-chargino 
mass region,  where the supersymmetric contribution has
a large destructive interference with the charged-Higgs boson 
contribution~\cite{Ciuchini:1998xy}.

There are also further analyses within the MFV hypothesis which try to include only the potentially 
large contributions beyond the leading order which are  enhanced 
by large $\tan\beta$ factors or by large logarithms of the form  $\ln(M_{\rm Susy}/M_W)$ where the masses of the supersymmmetric particles 
are assumed to be significantly larger than the $W$-boson mass~\cite{Degrassi:2000qf,Carena:2000uj, D'Ambrosio:2002ex}.  

\begin{figure}
\begin{center}
\epsfig{figure=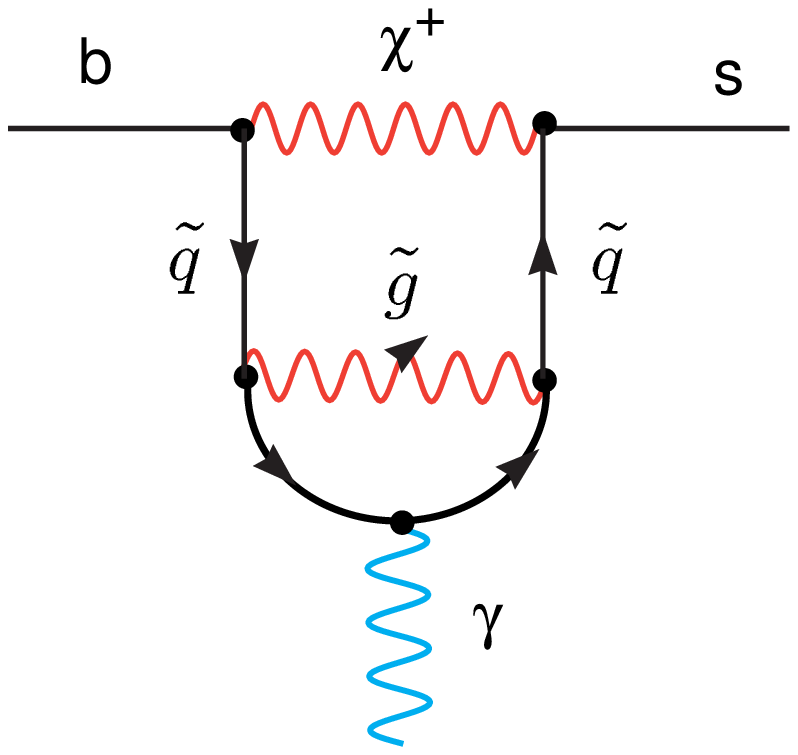,height=4.4cm} \epsfig{figure=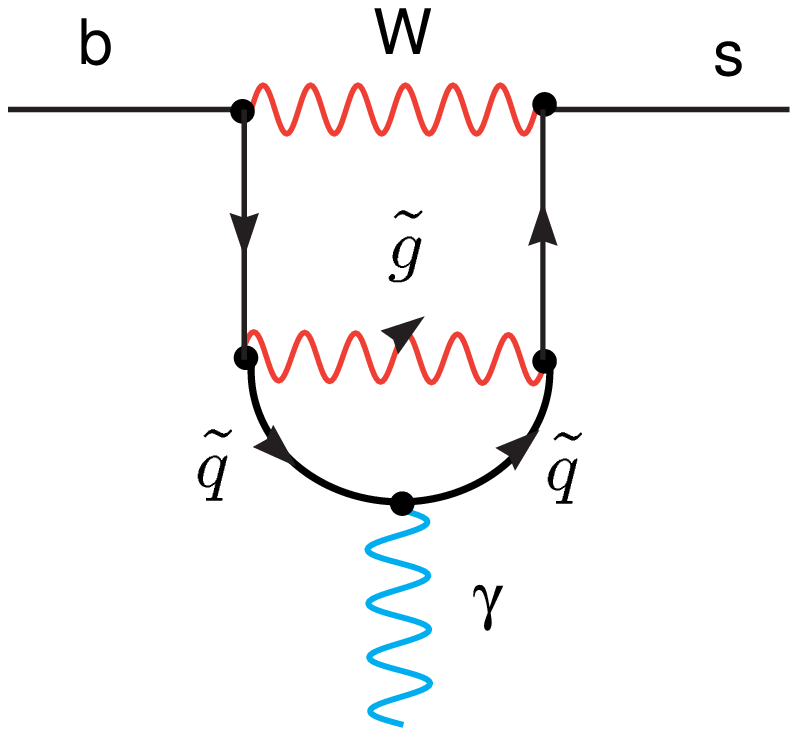,height=4.4cm} 
\end{center}
\caption{Example diagrams  for NLL gluino corrections to the chargino and W boson contribution.}
\label{Figure2}
\end{figure}

A practically  complete MFV analysis has been  presented in Ref.~\cite{Degrassi:2006eh}.
To  LL precison this calculation  includes  the  one-loop diagrams containing  a   $W$ boson  and  up-type quark, 
or  a charged Higgs boson and an up-type quark, or a chargino and an up-type squark (see Figure~\ref{Figure2}).
%Neutralino and gluino  exchange diagrams are  neglected under the  MFV assumption.
Neutralino and gluino exchange diagrams are neglected under the MFV assumption. 
To  NLL precision  the gluonic two-loop corrections to
the SM and charged Higgs loops are included, also 
two-loop diagrams with a
gluino together with a Higgs or W boson, and finally two-loop diagrams with a
chargino together with a gluon or a gluino or a quartic squark
coupling.  As already shown in Ref.\cite{Ciuchini:1998xy},  
the two-loop gluonic corrections to the chargino loops are not UV finite: in order to obtain a finite result one has to combine
them with the chargino-gluino diagrams.  

However, a MFV analysis  should take into account the fact that the simultaneous diagonalization of the quark
and squark mass matrices can be   imposed   at one scale $\mu_{\rm MFV}$. The renormalization group 
evolution  of the MSSM parameters then leads to a disalignement between the squark and 
quark mass matrices at scales different from $\mu_{\rm MFV}$~\cite{Degrassi:2006eh}.
So if the MFV condition is imposed at a scale   much larger than the superparticle mass scale $M_{\rm Susy}$,    very large logarithms of 
$M_{\rm Susy}/\mu_{\rm MFV}$  occur  in the Wilson coefficients. 
Then  the soft  Susy-breaking mass parameters -- which are assumed to be flavour-diagonal at the
scale $\mu_{\rm MFV}$ -- must be evolved down to $M_{\rm Susy}$  with the help of the 
appropriate renormalization group equations (RGE), thus,  generating some flavour violation in the squark
mass matrices which gets  absorbed in the couplings of the squark mass eigenstates with 
the gluinos and charginos. In Ref.~\cite{Degrassi:2006eh}, it is  argued, that the effects of the RGE-induced flavour mixing is relatively small  
and, therefore, are  only included  to  LL order, in the one-loop diagrams with gluinos and down-type squarks and in the one-loop diagrams 
with charginos and up-type squarks.  
There is a public computer code for this MFV calculation available which includes all contributions discussed above~\cite{Degrassi:2007kj}.

\subsection{NLL calculation in general MSSM}

Beyond minimal flavour violation, the most important role is played by the
non-diagonal gluino-quark-squark vertex due to the large strong coupling which
comes with this vertex. 
As discussed  above, this flavour non-diagonal vertex is induced by squark-mixing to the extent as it
is misaligned with quark mixing.  It represents  a  new flavour structure
beyond the SM Yukawa couplings. 
To understand these new  sources of flavour violation that may be present in
supersymmetric models in addition to those enclosed in the CKM matrix,
one has to consider the contributions to the squark mass matrices
\begin{equation}
{\cal M}_f^2 \equiv  \left( \begin{array}{cc}
  m^2_{\,f,\,LL} +F_{f\,LL} +D_{f\,LL}           & 
                 \left(m_{\,f,\,LR}^2\right) + F_{f\,LR} 
                                                     \\[1.01ex]
 \left(m_{\,f,\,LR}^{2}\right)^{\dagger} + F_{f\,RL} &
             \ \ m^2_{\,f,\,RR} + F_{f\,RR} +D_{f\,RR}                
 \end{array} \right) \,,
\label{squarku}
\end{equation}
where $f$ stands for up- or down-type squarks.
In the super-CKM basis, where the quark mass matrices are diagonal 
and the squarks are rotated in parallel to their superpartners,
the $F$ terms  from the superpotential and the $D$ terms 
from the gauge sector  turn out to be diagonal 
$3 \times 3$ submatrices of the 
$6 \times 6$
mass matrices ${\cal M}^2_f$. This is in general not true 
for the additional terms $m^2_{f, XY}$ with $X,Y \in \{L,R\}$, originating from  the soft 
supersymmetric breaking potential. 
Because the squark-quark-gluino coupling is flavour-diagonal 
in the super-CKM basis,   the gluino vertex in the  mass eigenstate
basis   is non-diagonal in flavour-space due to the off-diagonal
elements of the soft terms $m^2_{f,LL}$, $m^2_{f,RR}$, $m^2_{f,RL}$.

A complete LL analysis of the corresponding gluino contribution to the
inclusive decay rate of $\bar B \rightarrow X_s \gamma$ has been   
presented in Ref.~\cite{Borzumati:1999qt}. The sensitivity of the bounds on the down 
squark  mass matrix to radiative QCD LL corrections is  systematically analysed, 
including the SM and the gluino contributions. 
In Ref.~\cite{Besmer:2001cj} the interplay  between the various 
sources of flavour violation and the interference effects 
of SM, gluino,
chargino, neutralino and charged Higgs boson contributions is
 studied.  
The  bounds on  simple combinations of
elements of the soft part of the squark mass matrices
are found to be, in general, one order of magnitude weaker 
than the bound on the single off-diagonal element, which 
was derived in previous work  by
neglecting any kind of interference effects.  
Some effects beyond LL precision   like large $\tan \beta$ 
 effects are  estimated in Ref.~\cite{Okumura:2003hy} in analogy to the 
 MFV analyses  of  Refs.~\cite{Degrassi:2000qf,Carena:2000uj}.
 
\begin{figure}
\begin{center}
\epsfig{figure=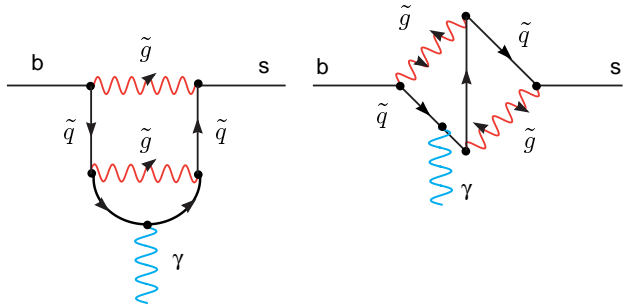,height=4.4cm} 
\end{center}
\caption{Example diagrams  for NLL gluino gluino contribution.}
\label{Figure3}
\end{figure}

Recently, the complete NLL  corrections 
to the Wilson coefficients (at the matching scale $\mu_W$) of the various versions of
magnetic and chromomagnetic operators which are induced by a
squark-gluino loop have been calculated~\cite{Greub:2011ji}. 
In this analysis  all the appearing heavy particles (which are the gluino, the
squarks and the top quark) are simultaneously integrated out at the high
scale. There are two classes of two-loop diagrams which have to be considered: 
diagrams with one gluino and a virtual gluon  and  diagrams with two gluinos (see Figure~\ref{Figure3}) or with one gluino and a squark-loop.
The former have been presented already in Ref.~\cite{Bobeth:1999ww} and now confirmed, while the latter 
have been calculated for the first time~\cite{Greub:2011ji}. 

Besides these NLL contributions due to the gluino vertex, 
there are of course more NLL corrections with non-minimal flavour violation;
they involve electroweak (gaugino and higgsino) vertices.
However, such contributions are in general suppressed compared to the ones related to the 
gluino.
There are two types of such contributions at the NLL level:  First, there are
electroweak corrections to the non-minimal LL gluino contribution (in which 
the electroweak vertex is flavour-diagonal or MFV-like) which are naturally
suppressed due to the smaller coupling constants and due to the CKM hierarchy. 
Second, there is also non-minimal flavour violation via squark-mixing 
in the electroweak vertices possible. 
But such contributions are already
 suppressed at the LL level compared to the gluino contribution  
due to the smaller  coupling constant, apart from the chargino contributions in specific  parts of the parameter space 
in which for example the trilinear coupling $A^u_{23}$ is  very large. 
These features  do not change of course when gluon- and gluino-induced  NLL corrections are added 
to such  LL contributions. Still, the leading chirally enhanced corrections can be
easily calculated by 
inserting the effective Feynman rules of \cite{Crivellin:2011jt}
into the results of \cite{Besmer:2001cj}.
Summing up,  the complete NLL corrections induced by the gluino vertex given in Ref.~\cite{Greub:2011ji}  represent the dominant 
contribution beyond MFV at this order in most parts of the MSSM parameter space. 
They  are complementary to the MFV contributions at the NLL level which are given  in Ref.~\cite{Degrassi:2006eh}.
The results are presented also in public  computer code~\cite{Greub:2011ji}.

\section*{Acknowledgments}
TH thanks the organizers of the conference  for the interesting and valuable meeting  and the CERN theory group for its  hospitality during his regular visits to CERN where part of this work  was written. \\


\begin{thebibliography}{12}

%%%%%%%Introduction 


\bibitem{Hurth:2010tk}
  T.~Hurth and M.~Nakao,
  %``Radiative and Electroweak Penguin Decays of B Mesons,''
  Ann.\ Rev.\ Nucl.\ Part.\ Sci.\  {\bf 60}, 645 (2010)
  [arXiv:1005.1224 [hep-ph]].
  %%CITATION = ARNUA,60,645;%%


\bibitem{Hurth:2007xa}
  T.~Hurth,
  %``Status of SM calculations of b > s transitions,''
  Int.\ J.\ Mod.\ Phys.\  A {\bf 22} (2007) 1781
  [arXiv:hep-ph/0703226].
  %%CITATION = IMPAE,A22,1781;%%




\bibitem{Hurth:2003vb}
  T.~Hurth,
  %``Present status of inclusive rare B decays,''
  Rev.\ Mod.\ Phys.\  {\bf 75}, 1159 (2003)
  [arXiv:hep-ph/0212304].
  %%CITATION = RMPHA,75,1159;%%
%\cite{Ghinculov:2002ge}



\bibitem{Benzke:2010js}
  M.~Benzke, S.~J.~Lee, M.~Neubert and G.~Paz,
  %``Factorization at Subleading Power and Irreducible Uncertainties in $\bar
  %B\to X_s\gamma$ Decay,''
  JHEP {\bf 1008}, 099 (2010)
  [arXiv:1003.5012 [hep-ph]].
  %%CITATION = JHEPA,1008,099;%%



\bibitem{Misiak:2006zs}
  M.~Misiak {\it et al.},
  %``The first estimate of B(anti-B --> X/s gamma) at O(alpha(s)**2),''
  Phys.\ Rev.\ Lett.\  {\bf 98}, 022002 (2007)
  [arXiv:hep-ph/0609232].
  %%CITATION = PRLTA,98,022002;%%
 

  

\bibitem{Misiak:2004ew}
  M.~Misiak and M.~Steinhauser,
  %``Three loop matching of the dipole operators for b ---> s gamma and b ---> s
  %g,''
  Nucl.\ Phys.\  B {\bf 683} (2004) 277
  [arXiv:hep-ph/0401041].
  %%CITATION = NUPHA,B683,277;%%


\bibitem{Bobeth:1999mk}
  C.~Bobeth, M.~Misiak and J.~Urban,
  %``Photonic penguins at two loops and m(t) dependence of BR[B ---> X(s)
  %lepton+ lepton-],''
  Nucl.\ Phys.\  B {\bf 574}, 291 (2000)
  [arXiv:hep-ph/9910220].
  %%CITATION = NUPHA,B574,291;%%


\bibitem{Gorbahn:2004my}
  M.~Gorbahn and U.~Haisch,
  %``Effective Hamiltonian for non-leptonic |Delta F| = 1 decays at NNLO in
  %QCD,''
  Nucl.\ Phys.\  B {\bf 713}, 291 (2005)
  [arXiv:hep-ph/0411071].
  %%CITATION = NUPHA,B713,291;%%


\bibitem{Gorbahn:2005sa}
  M.~Gorbahn, U.~Haisch and M.~Misiak,
  %``Three-loop mixing of dipole operators,''
  Phys.\ Rev.\ Lett.\  {\bf 95}, 102004 (2005)
  [arXiv:hep-ph/0504194].
  %%CITATION = PRLTA,95,102004;%%


\bibitem{Czakon:2006ss}
  M.~Czakon, U.~Haisch and M.~Misiak,
  %``Four-Loop Anomalous Dimensions for Radiative Flavour-Changing Decays,''
  JHEP {\bf 0703}, 008 (2007)
  [arXiv:hep-ph/0612329].
  %%CITATION = JHEPA,0703,008;%%


\bibitem{Blokland:2005uk}
  I.~R.~Blokland, A.~Czarnecki, M.~Misiak, M.~Slusarczyk and F.~Tkachov,
  %``The Electromagnetic dipole operator effect on anti-B ---> X(s) gamma at
  %O(alpha**2(s)),''
  Phys.\ Rev.\  D {\bf 72}, 033014 (2005)
  [arXiv:hep-ph/0506055].
  %%CITATION = PHRVA,D72,033014;%%


\bibitem{Melnikov:2005bx}
  K.~Melnikov and A.~Mitov,
  %``The Photon energy spectrum in B ---> X(s) + gamma in perturbative QCD
  %through O(alpha(s)**2),''
  Phys.\ Lett.\  B {\bf 620}, 69 (2005)
  [arXiv:hep-ph/0505097].
  %%CITATION = PHLTA,B620,69;%%



\bibitem{Asatrian:2006ph}
  H.~M.~Asatrian, A.~Hovhannisyan, V.~Poghosyan, T.~Ewerth, C.~Greub and T.~Hurth,
  %``NNLL QCD contribution of the electromagnetic dipole operator to
  %Gamma(anti-B ---> X(s) gamma),''
  Nucl.\ Phys.\  B {\bf 749}, 325 (2006)
  [arXiv:hep-ph/0605009].
  %%CITATION = NUPHA,B749,325;%%


\bibitem{Asatrian:2006sm}
  H.~M.~Asatrian, T.~Ewerth, A.~Ferroglia, P.~Gambino and C.~Greub,
  %``Magnetic dipole operator contributions to the photon energy spectrum in
  %anti-B ---> X(s) gamma at O(alpha**2(s)),''
  Nucl.\ Phys.\  B {\bf 762}, 212 (2007)
  [arXiv:hep-ph/0607316].
  %%CITATION = NUPHA,B762,212;%%


\bibitem{Bieri:2003ue}
  K.~Bieri, C.~Greub and M.~Steinhauser,
  %``Fermionic NNLL corrections to b ---> s gamma,''
  Phys.\ Rev.\  D {\bf 67}, 114019 (2003)
  [arXiv:hep-ph/0302051].
  %%CITATION = PHRVA,D67,114019;%%


\bibitem{Misiak:2006ab}
  M.~Misiak and M.~Steinhauser,
  %``NNLO QCD corrections to the B -> X_s gamma matrix elements using
  %interpolation in m_c,''
  Nucl.\ Phys.\  B {\bf 764} (2007) 62
  [arXiv:hep-ph/0609241].
  %%CITATION = NUPHA,B764,62;%%

\bibitem{hfag}
  D.~Asner {\it et al.}  [Heavy Flavor Averaging Group],
  %``Averages of b-hadron, c-hadron, and $\tau-lepton Properties,''
  arXiv:1010.1589 [hep-ex] \\ and online update at
http://www.slac.stanford.edu/xorg/hfag (2010)











\bibitem{Antonelli:2009ws}
  M.~Antonelli {\it et al.},
  %``Flavor Physics in the Quark Sector,''
  Phys.\ Rept.\  {\bf 494} (2010) 197
  [arXiv:0907.5386 [hep-ph]].
  %%CITATION = PRPLC,494,197;%%



\bibitem{Buchalla:2008jp}
  M.~Artuso {\it et al.},
  %``$B$, $D$ and $K$ decays,''
  Eur.\ Phys.\ J.\  C {\bf 57}, 309 (2008)
  [arXiv:0801.1833 [hep-ph]].
  %%CITATION = EPHJA,C57,309;%%



\bibitem{Belle} 
  Belle collaboration: http://belle.kek.jp/

\bibitem{Babar} 
  BaBar collaboration: http://www.slac.stanford.edu/BFROOT/

\bibitem{TevatronB1}
  CDF collaboration: http://www-cdf.fnal.gov/physics/new/bottom/bottom.html

\bibitem{TevatronB2}
  D0 collaboration: http://www-d0.fnal.gov/Run2Physics/WWW/results/b.htm



\bibitem{LHCb}  G.~Raven, {\it B Physics Results from the LHC}, 
talk at Lepton-Photon 2011, 22--27 Aug.\ 2011, %Tata Institute of Fundamental Research, 
Mumbai, India.

%%%%%%%  Chapter 2


\bibitem{Chivukula:1987py}
  R.~S.~Chivukula and H.~Georgi,
  %``Composite Technicolor Standard Model,''
  Phys.\ Lett.\ B {\bf 188} (1987) 99.
  %%CITATION = PHLTA,B188,99;%%



\bibitem{Hall:1990ac}
  L.~J.~Hall and L.~Randall,
  %``Weak scale effective supersymmetry,''
  Phys.\ Rev.\ Lett.\  {\bf 65}, 2939 (1990).
  %%CITATION = PRLTA,65,2939;%%


\bibitem{D'Ambrosio:2002ex}
  G.~D'Ambrosio, G.~F.~Giudice, G.~Isidori and A.~Strumia,
  %``Minimal flavour violation: An effective field theory approach,''
  Nucl.\ Phys.\  B {\bf 645}, 155 (2002)
  [arXiv:hep-ph/0207036].
  %%CITATION = NUPHA,B645,155;%%


\bibitem{Hurth:2009ke}
  T.~Hurth and W.~Porod,
  %``Flavour violating squark and gluino decays,''
  JHEP {\bf 0908}, 087 (2009)
  [arXiv:0904.4574 [hep-ph]].
  %%CITATION = JHEPA,0908,087;%%
%% Wilson coefficients at low scale 




\bibitem{Ferrara}
S.~Ferrara and E.~Remiddi,
%``Absence of the anomalous magnetic moment in a supersymmetric abelian gauge 
%theory,''
Phys.\ Lett.\ B {\bf 53}, 347 (1974).



%%%%%%%%%%%%%%



%%%MFV 

%\cite{Haisch:2007vb}
\bibitem{Haisch:2007vb}
  U.~Haisch and A.~Weiler,
  %``Bound on minimal universal extra dimensions from anti-B --> X/s gamma,''
  Phys.\ Rev.\  D {\bf 76}, 034014 (2007)
  [arXiv:hep-ph/0703064].
  %%CITATION = PHRVA,D76,034014;%%

%\cite{Ciuchini:1997xe}
\bibitem{Ciuchini:1997xe}
  M.~Ciuchini, G.~Degrassi, P.~Gambino and G.~F.~Giudice,
  %``Next-to-leading QCD corrections to B --> X/s gamma: Standard model and
  %two-Higgs doublet model,''
  Nucl.\ Phys.\  B {\bf 527} (1998) 21
  [arXiv:hep-ph/9710335].
  %%CITATION = NUPHA,B527,21;%%

%\cite{Borzumati:1998tg}
\bibitem{Borzumati:1998tg}
  F.~Borzumati and C.~Greub,
  %``2HDMs predictions for anti-B --> X/s gamma in NLO {QCD},''
  Phys.\ Rev.\  D {\bf 58}, 074004 (1998)
  [arXiv:hep-ph/9802391].
  %%CITATION = PHRVA,D58,074004;%%

  
  
  
  
%\cite{Bertolini:1990if}
\bibitem{Bertolini:1990if}
  S.~Bertolini, F.~Borzumati, A.~Masiero and G.~Ridolfi,
  %``Effects of supergravity induced electroweak breaking on rare $B$ decays and
  %mixings,''
  Nucl.\ Phys.\  B {\bf 353}, 591 (1991).
  %%CITATION = NUPHA,B353,591;%%


%\cite{Ciuchini:1998xy}
\bibitem{Ciuchini:1998xy}
  M.~Ciuchini, G.~Degrassi, P.~Gambino, G.~F.~Giudice,
  %``Next-to-leading QCD corrections to B ---> X(s) gamma in supersymmetry,''
  Nucl.\ Phys.\  {\bf B534}, 3-20 (1998).
  [hep-ph/9806308].


%\cite{Degrassi:2000qf}
\bibitem{Degrassi:2000qf}
  G.~Degrassi, P.~Gambino and G.~F.~Giudice,
  %``B --> X/s gamma in supersymmetry: Large contributions beyond the  leading
  %order,''
  JHEP {\bf 0012}, 009 (2000)
  [arXiv:hep-ph/0009337].
  %%CITATION = JHEPA,0012,009;%%

%\cite{Carena:2000uj}
\bibitem{Carena:2000uj}
  M.~S.~Carena, D.~Garcia, U.~Nierste and C.~E.~M.~Wagner,
  %``b --> s gamma and supersymmetry with large tan(beta),''
  Phys.\ Lett.\  B {\bf 499}, 141 (2001)
  [arXiv:hep-ph/0010003].
  %%CITATION = PHLTA,B499,141;%%



%\cite{Borzumati:1999qt}
\bibitem{Borzumati:1999qt}
  F.~Borzumati, C.~Greub, T.~Hurth and D.~Wyler,
  %``Gluino contribution to radiative B decays: Organization of QCD  corrections
  %and leading order results,''
  Phys.\ Rev.\  D {\bf 62}, 075005 (2000)
  [arXiv:hep-ph/9911245].
  %%CITATION = PHRVA,D62,075005;%%

%\cite{Besmer:2001cj}
\bibitem{Besmer:2001cj}
  T.~Besmer, C.~Greub and T.~Hurth,
  %``Bounds on flavor violating parameters in supersymmetry,''
  Nucl.\ Phys.\  B {\bf 609}, 359 (2001)
  [arXiv:hep-ph/0105292].
  %%CITATION = NUPHA,B609,359;%%

%\cite{Ciuchini:2002uv}
\bibitem{Ciuchini:2002uv}
  M.~Ciuchini, E.~Franco, A.~Masiero and L.~Silvestrini,
  %``b --> s transitions: A new frontier for indirect SUSY searches,''
  Phys.\ Rev.\  D {\bf 67}, 075016 (2003)
  [Erratum-ibid.\  D {\bf 68}, 079901 (2003)]
  [arXiv:hep-ph/0212397].
  %%CITATION = PHRVA,D67,075016;%%

%\cite{Ciuchini:2003rg}
\bibitem{Ciuchini:2003rg}
  M.~Ciuchini, A.~Masiero, L.~Silvestrini, S.~K.~Vempati and O.~Vives,
  %``Grand unification of quark and lepton FCNCs,''
  Phys.\ Rev.\ Lett.\  {\bf 92}, 071801 (2004)
  [arXiv:hep-ph/0307191].
  %%CITATION = PRLTA,92,071801;%%
  
  
  %\cite{Okumura:2003hy}
\bibitem{Okumura:2003hy}
  K.~i.~Okumura and L.~Roszkowski,
  %``Large beyond-leading-order effects in b --> s gamma in supersymmetry  with
  %general flavor mixing,''
  JHEP {\bf 0310}, 024 (2003)
  [arXiv:hep-ph/0308102].
  %%CITATION = JHEPA,0310,024;


  
  
  %\cite{Degrassi:2006eh}
\bibitem{Degrassi:2006eh}
  G.~Degrassi, P.~Gambino and P.~Slavich,
  %``QCD corrections to radiative B decays in the MSSM with minimal flavor
  %violation,''
  Phys.\ Lett.\  B {\bf 635}, 335 (2006)
  [arXiv:hep-ph/0601135].
  %%CITATION = PHLTA,B
  



  
%\cite{Ciuchini:2007ha}
\bibitem{Ciuchini:2007ha}
  M.~Ciuchini, A.~Masiero, P.~Paradisi, L.~Silvestrini, S.~K.~Vempati and O.~Vives,
  %``Soft SUSY breaking grand unification: Leptons versus quarks on the flavor
  %playground,''
  Nucl.\ Phys.\  B {\bf 783}, 112 (2007)
  [arXiv:hep-ph/0702144].
  %%CITATION = NUPHA,B783,112;%%
  
%\cite{Altmannshofer:2008vr}
\bibitem{Altmannshofer:2008vr}
  W.~Altmannshofer, D.~Guadagnoli, S.~Raby and D.~M.~Straub,
  %``SUSY GUTs with Yukawa unification: A Go/no-go study using FCNC processes,''
  Phys.\ Lett.\  B {\bf 668}, 385 (2008)
  [arXiv:0801.4363 [hep-ph]].
  %%CITATION = PHLTA,B668,385;%%





%\cite{Crivellin:2009ar}
\bibitem{Crivellin:2009ar}
  A.~Crivellin and U.~Nierste,
  %``Chirally enhanced corrections to FCNC processes in the generic MSSM,''
  Phys.\ Rev.\  D {\bf 81}, 095007 (2010)
  [arXiv:0908.4404 [hep-ph]].
  %%CITATION = PHRVA,D81,095007;%%



%\cite{Crivellin:2011jt}
\bibitem{Crivellin:2011jt}
  A.~Crivellin, L.~Hofer, J.~Rosiek,
  %``Complete resummation of chirally-enhanced loop-effects in the MSSM with non-minimal sources of flavor-violation,''
  JHEP {\bf 1107}, 017 (2011).
  [arXiv:1103.4272 [hep-ph]].


  

%\cite{Ali:2002jg}
\bibitem{Ali:2002jg}
  A.~Ali, E.~Lunghi, C.~Greub and G.~Hiller,
  %``Improved model independent analysis of semileptonic and radiative rare $B$
  %decays,''
  Phys.\ Rev.\  D {\bf 66}, 034002 (2002)
  [arXiv:hep-ph/0112300].
  %%CITATION = PHRVA,D66,034002;%%


%\cite{Hurth:2008jc}
\bibitem{Hurth:2008jc}
  T.~Hurth, G.~Isidori, J.~F.~Kamenik and F.~Mescia,
  %``Constraints on New Physics in MFV models: A Model-independent analysis of
  %$\Delta$ F = 1 processes,''
  Nucl.\ Phys.\  B {\bf 808} (2009) 326
  [arXiv:0807.5039 [hep-ph]].
  %%CITATION = NUPHA,B808,326;%%






%%%%%%%%%%%%%%%%%%%%%%%%%%%%%

%\cite{Degrassi:2007kj}
\bibitem{Degrassi:2007kj}
  G.~Degrassi, P.~Gambino and P.~Slavich,
  %``SusyBSG: a fortran code for BR[B -> Xs gamma] in the MSSM with Minimal
  %Flavor Violation,''
  Comput.\ Phys.\ Commun.\  {\bf 179}, 759 (2008)
  [arXiv:0712.3265 [hep-ph]].
  %%CITATION = CPHCB,179,759;%%


%\cite{Greub:2011ji}
\bibitem{Greub:2011ji}
  C.~Greub, T.~Hurth, V.~Pilipp, C.~Schupbach, M.~Steinhauser,
  %``Complete next-to-leading order gluino contributions to b--> s gamma and b--> s g,''
  Nucl.\ Phys.\  {\bf B853}, 240-276 (2011).
  [arXiv:1105.1330 [hep-ph]].



%\cite{Bobeth:1999ww}
\bibitem{Bobeth:1999ww}
  C.~Bobeth, M.~Misiak, J.~Urban,
  %``Matching conditions for b ---> s gamma and b ---> s gluon in extensions of the standard model,''
  Nucl.\ Phys.\  {\bf B567}, 153-185 (2000).
  [hep-ph/9904413].


\end{thebibliography}
\end{document}